\documentclass[conference]{IEEEtran}
\IEEEoverridecommandlockouts

\usepackage{amsmath,graphicx}
\usepackage{graphicx}
\usepackage{caption}
\usepackage{subcaption}
\usepackage{algorithm}
\usepackage{algorithmic}

\usepackage{amsmath,amssymb}
\usepackage[keeplastbox]{flushend}
\usepackage{balance}

\usepackage{cite}
\usepackage{amsmath,amssymb,amsfonts}
\usepackage{algorithmic}
\usepackage{graphicx}
\usepackage{textcomp}
\usepackage{xcolor}
\def\BibTeX{{\rm B\kern-.05em{\sc i\kern-.025em b}\kern-.08em
    T\kern-.1667em\lower.7ex\hbox{E}\kern-.125emX}}
\begin{document}

\title{Online unsupervised deep unfolding \\ for MIMO channel estimation
}

\author{\IEEEauthorblockN{Luc Le Magoarou, St\'ephane Paquelet}
\IEEEauthorblockA{
\textit{b$<>$com}\\
Rennes, France}
}

\maketitle

\begin{abstract}
Channel estimation is a difficult problem in MIMO systems. Using a physical model allows to ease the problem, injecting a priori information based on the physics of propagation. However, such models rest on simplifying assumptions and require to know precisely the system configuration, which is unrealistic.
In this paper, we propose to perform online learning for channel estimation in a massive MIMO context, adding flexibility to physical models by unfolding a channel estimation algorithm (matching pursuit) as a neural network. This leads to a computationally efficient neural network that can be trained online when initialized with an imperfect model. The method allows a base station to automatically correct its channel estimation algorithm based on incoming data, without the need for a separate offline training phase.
It is applied to realistic channels and shows great performance, achieving channel estimation error almost as low as one would get with a perfectly calibrated system.
\end{abstract}

\begin{IEEEkeywords}
Autoencoders, deep unfolding, MIMO channel estimation, online learning.
\end{IEEEkeywords}

\section{Introduction}

Data processing techniques are often based on the \emph{manifold assumption}: Meaningful data (signals) lie near a low dimensional manifold, although their apparent dimension is much larger \cite{Carlsson2009,Peyre2009}.

In MIMO channel estimation, using a physical model amounts to parameterize a manifold by physical parameters such as the directions, delays and gains of the propagation paths, the dimension of the manifold being equal to the number of real parameters considered in the model. Physical models allow to inject strong a priori knowledge based on solid principles \cite{Sayeed2002,Lemagoarou2018}, but necessarily make simplifying assumptions (e.g. the plane wave assumption \cite{Lemagoarou2019b}) and require exact knowledge of the system configuration (positions of the antennas, gains, etc.).

On the other hand, machine learning techniques have recently led to tremendous successes in various domains \cite{Lecun2015}. Their main feature is to learn the data representation (manifold) directly on training data, without requiring any specific a priori knowledge. This flexibility in the manifold construction comes at the price of computationally heavy learning and difficulties to inject knowledge on the problem at hand.

Recently, it has been proposed to unfold iterative inference algorithms so as to express them as neural networks that can be optimized \cite{Gregor2010,Hershey2014}. This has the advantage of adding flexibility to algorithms based on classical models, and amounts to constrain the search for the appropriate manifold with a priori knowledge on the problem at hand. Moreover, this leads to inference algorithms of reduced complexity \cite{Monga2019}.

\noindent{\bf Contributions.} In this letter, we propose to perform \emph{online learning} for channel estimation in a massive MIMO context. Starting from an imperfect physical channel model, our method allows a base station to automatically correct its channel estimation algorithm based on incoming data, without the need for a separate offline training phase. It is based on the unfolding of the matching pursuit algorithm, which is simple and computationally efficient. The obtained neural network is trained in an \emph{unsupervised} way. The overall complexity of the forward and backward passes in the network is of the same order as performing channel estimation only (without any learning), which makes online learning feasible. Such a method is particularly suited to imperfectly known or non-calibrated systems. Note that since this paper has been written, we further developed the introduced ideas in a longer paper \cite{Yassine2020} (still in a preprint state, not submitted anywhere). This longer paper introduces an automatic adaptation to the signal to noise ratio (SNR) and demonstrates several potential applications of the method.

\noindent{\bf Related work.} Machine learning holds promise for wireless communications (see \cite{Oshea2017,Wang2017} for exhaustive surveys). It has recently been proposed to use adaptive data representations for MIMO channel estimation using dictionary learning techniques \cite{Ding2018}. However, dictionary learning with algorithms such as K-SVD \cite{Aharon2006} as proposed in \cite{Ding2018} is very computationally heavy, and thus not suited to online learning.

Deep unfolding has also been considered by communication researchers (see \cite{Balatsoukas2019} and references therein). It has been proposed in \cite{He2018} to perform channel estimation in a massive MIMO context, based on the unfolding of a sparse recovery algorithm (namely denoising-based approximate message passing \cite{Metzler2016}). However, the method is directly adapted from image processing and does not make use of a physical channel model as initialization. A recent work also proposes to use deep unfolding for channel estimation \cite{Wei2019}, but using a physical model to optimize the shrinkage function. However, previously proposed methods based on unfolding all require an offline training phase and are of high complexity  compared to classical methods \cite{Vanlier2020}.

The main novelty of this letter is the online nature of the method, that does not require a separate offline learning phase, since learning is done while using the channel estimation algorithm that corrects itself over time. This is made possible by the very low complexity of the considered estimation algorithm.


\section{Problem formulation}

\noindent {\bf System settings.} We consider in this letter a massive MIMO system, also known as multi-user MIMO (MU-MIMO) system \cite{Rusek2013}, in which a base station equipped with $N$ antennas communicates with $K$ single antenna users ($K<N$). The system operates in time division duplex (TDD) mode, so that channel reciprocity holds and the channel is estimated in the uplink: each user sends a pilot sequence $\mathbf{p}_k$ (orthogonal to the sequences of the other users, $\mathbf{p}_k^H\mathbf{p}_l = \delta_{kl}$) for the base station to estimate the channel. The received signal is thus expressed $\mathbf{R}=\sum_{k=1}^K\mathbf{h}_k\mathbf{p}_k^H +\mathbf{N}$, where $\mathbf{N}$ is noise.
After correlating the received signal with the pilot sequences, and assuming no pilot contamination from adjacent cells for simplicity, the base station gets noisy measurements of the channels of all users, each taking the canonical form 
\begin{equation}
\mathbf{x} = \mathbf{h} + \mathbf{n},
\label{eq:observations}
\end{equation}
where $\mathbf{h}$ is the channel of the considered user and $\mathbf{n}$ is the noise, with $\mathbf{n} \sim \mathcal{CN}(0,\sigma^2\mathbf{Id})$. We drop the user index $k$ here and in the following, since our approach treats the channels of all users the same way. Note that $\mathbf{x}$ is already an unbiased estimator of the channel, and we call it the least squares (LS) estimator in the sequel. Its performance can be assessed by the signal to noise ratio (SNR)
$$
\text{SNR}_{\text{in}}  \triangleq \frac{\left\Vert \mathbf{h} \right\Vert_2^2}{N\sigma^2}.
$$ 
However, one can get better channel estimates using a physical model, as is explained in the next paragraph.

\noindent {\bf Physical model.}Let us denote $\{g_1,\dots,g_N\}$ the complex gains of the base stations antennas and $\{\overrightarrow{a_1},\dots,\overrightarrow{a_N}\}$ their positions with respect to the centroid of the antenna array. Then, under the plane wave assumption and assuming omnidirectional antennas (isotropic radiation patterns), the channel resulting from a single propagation path with direction of arrival (DoA) $\overrightarrow{u}$ is proportional to the \emph{steering vector}
$$
\mathbf{e}(\overrightarrow{u}) \triangleq (
g_1\mathrm{e}^{-\mathrm{j}\frac{2\pi}{\lambda}\overrightarrow{a_1}.\overrightarrow{u}}, \dots, g_N\mathrm{e}^{-\mathrm{j}\frac{2\pi}{\lambda}\overrightarrow{a_N}.\overrightarrow{u}})^T 
$$
which reads $\mathbf{h} = \beta \mathbf{e}(\overrightarrow{u}),$
with $\beta \in \mathbb{C}$. In that case, a sensible estimation strategy \cite{Sayeed2002,Lemagoarou2018} is to build a dictionary of steering vectors corresponding to $A$ potential DoAs: $\mathbf{E} \triangleq \begin{pmatrix}
\mathbf{e}(\overrightarrow{u_1}),\dots,\mathbf{e}(\overrightarrow{u_A})
\end{pmatrix}
$ and to compute a channel estimate with the procedure
\begin{align}
\begin{split}
&\overrightarrow{v}=\text{argmax}_{\overrightarrow{u_i}} \,\,\, |\mathbf{e}(\overrightarrow{u_i})^H\mathbf{x}|, \\
&\hat{\mathbf{h}} = \mathbf{e}(\overrightarrow{v})\mathbf{e}(\overrightarrow{v})^H\mathbf{x}.
\end{split}
\label{eq:estim_strat}
\end{align} 
The first step of this procedure amounts to find in the dictionary the most correlated column with the observation to estimate the DoA and the second step amounts to project the observation on the corresponding steering vector. The SNR at the output of this procedure reads
$$\text{SNR}_\text{out} \triangleq \frac{\Vert \mathbf{h} \Vert_2^2}{\mathbb{E}\big[\Vert\mathbf{h}-\hat{\mathbf{h}}\Vert_2^2\big]},$$
and we have at best $\text{SNR}_\text{out} =N\text{SNR}_\text{in}$ (neglecting the discretization error).
 
Note that the evoked strategy can be generalized to multipath channels of the form
$
\mathbf{h} = \sum\nolimits_{p=1}^P\beta_p \mathbf{e}(\overrightarrow{u_p}),
$
using greedy sparse recovery algorithms such as matching pursuit (MP) or orthogonal matching pursuit (OMP) \cite{Mallat1993}.

\begin{figure}[tbp]
    \centering
    \includegraphics[width=0.95\columnwidth]{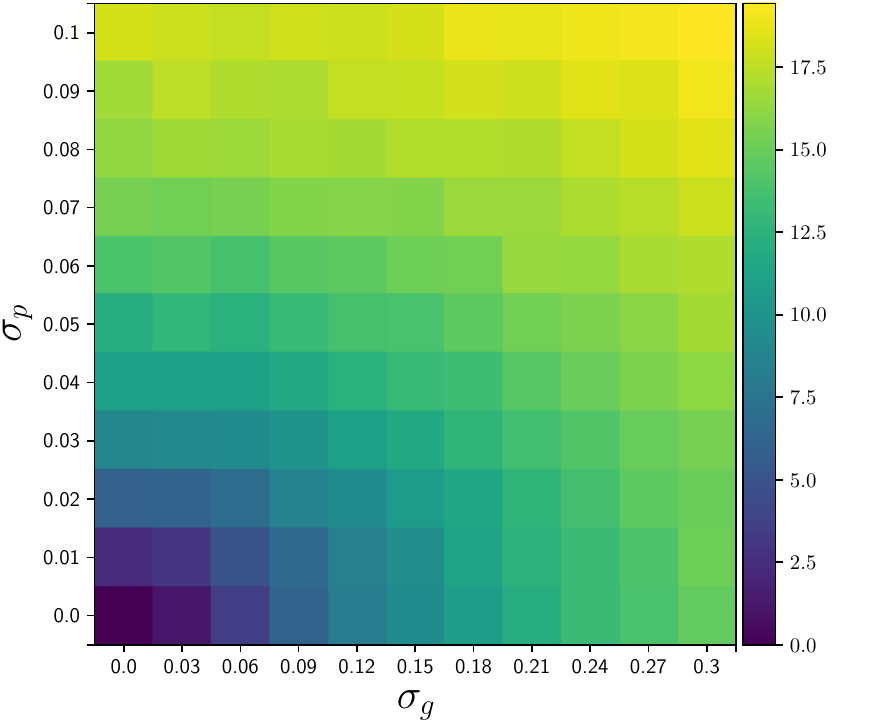}    
    \caption{SNR loss in decibels (dB) due to imperfect knowledge of the system.}
    \label{fig:imperfect_model} 
\end{figure}

\section{Impact of imperfect models}
\label{sec:impact}
The estimation strategy based on a physical model requires knowing the system configuration (antenna gains and positions) and necessarily relies on hypotheses. What happens if the configuration is imperfectly known or if some hypotheses are not valid? In order to answer this simple question, let us perform an experiment. 
Consider an antenna array of $N=64$ antennas at the base station, whose known \emph{nominal} configuration is an uniform linear array (ULA) of unit gain antennas separated by half-wavelengths and aligned with the $x$-axis. This nominal configuration corresponds to gains and positions $\{\tilde{g}_i,\tilde{\overrightarrow{a_i}}\}_{i=1}^N$. Now, suppose the knowledge of the system configuration is imperfect, meaning that the unknown \emph{true} configuration of the system is given by the gains and positions $\{g_i,\overrightarrow{a_i}\}_{i=1}^N$, with 
\begin{align}
\begin{split}
&g_i = \tilde{g}_i + n_{g,i}, \, n_{g,i} \sim \mathcal{CN}(0,\sigma_g^2),\\
&\overrightarrow{a_i} = \tilde{\overrightarrow{a_i}} + \lambda\mathbf{n}_{p,i}, \, \mathbf{n}_{p,i} = {\small\begin{pmatrix} e_{p,i}, &0, & 0 \end{pmatrix}^T}, e_{p,i} \sim \mathcal{N}(0,\sigma_p^2).
\end{split}
\label{eq:imperfection}
\end{align}
This way, $\sigma_g$ (resp. $\sigma_p$) quantifies the uncertainty about the antenna gains (resp. spacings). Moreover, let 
$$
\tilde{\mathbf{e}}(\overrightarrow{u}) \triangleq (
\tilde{g}_1\mathrm{e}^{-\mathrm{j}\frac{2\pi}{\lambda}\tilde{\overrightarrow{a_1}}.\overrightarrow{u}},\dots,
\tilde{g}_N\mathrm{e}^{-\mathrm{j}\frac{2\pi}{\lambda}\tilde{\overrightarrow{a_N}}.\overrightarrow{u}}
)^T
$$
be the nominal steering vector and $\tilde{\mathbf{E}} \triangleq \begin{pmatrix}
\tilde{\mathbf{e}}(\overrightarrow{u_1}),\dots,\tilde{\mathbf{e}}(\overrightarrow{u_A})
\end{pmatrix}$ be a dictionary of nominal steering vectors. The experiment consists in comparing the estimation strategy of \eqref{eq:estim_strat} using the true (perfect but unknown) dictionary $\mathbf{E}$ with the exact same strategy using the nominal (imperfect but known) dictionary $\tilde{\mathbf{E}}$. To do so, we generate measurements according to \eqref{eq:observations} with channels of the form $\mathbf{h} = \mathbf{e}(\overrightarrow{u})$ where $\overrightarrow{u}$ corresponds to azimuth angles chosen uniformly at random, and $\text{SNR}_{\text{in}}$ is set to $10\,\text{dB}$. Then, the dictionaries $\mathbf{E}$ and $\tilde{\mathbf{E}}$ are built by choosing $A=32N$ directions corresponding to evenly spaced azimuth angles. Let $\hat{\mathbf{h}}_{\mathbf{E}}$ be the estimate obtained using $\mathbf{E}$ in \eqref{eq:estim_strat}, and $\hat{\mathbf{h}}_{\tilde{\mathbf{E}}}$ the estimate obtained using $\tilde{\mathbf{E}}$. The SNR loss caused by using $\tilde{\mathbf{E}}$ instead of $\mathbf{E}$ is measured by the quantity $\Vert \hat{\mathbf{h}}_{\tilde{\mathbf{E}}} - \mathbf{h} \Vert_2^2/\Vert \hat{\mathbf{h}}_\mathbf{E} - \mathbf{h} \Vert_2^2$. Results in terms of SNR loss, in average over $10$ antenna array realizations and $1000$ channel realizations per antenna array realization are shown on figure~\ref{fig:imperfect_model}. From the figure, it is obvious that even a relatively small uncertainty about the system configuration can cause a great SNR loss. For example, an uncertainty of $0.03\lambda$ on the antenna spacings and of $0.09$ on the antenna gains leads to an SNR loss of more than $10\,\text{dB}$, which means that the mean squared error is increased more than ten times. This experiment highlights the fact that using imperfect models can severely harm estimation performance. The main contribution of this letter is to propose a way to correct imperfect physical models using machine learning.

\section{Deep unfolding strategy}
Let us now propose a strategy based on deep unfolding allowing to correct a channel estimation algorithm based on an imperfect physical model incrementally, via online learning.

 \subsection{Basic principle}

\noindent{\bf Unfolding.} The estimation strategy of \eqref{eq:estim_strat} can be unfolded as a neural network taking the observation $\mathbf{x}$ as input and outputting a channel estimate $\hat{\mathbf{h}}$. Indeed, the first step amounts to perform a linear transformation (multiplying the input by the matrix $\mathbf{E}^H$) followed by a nonlinear one (finding the inner product of maximum amplitude and setting all the others to zero) and the second step corresponds to a linear transformation (multiplying by the matrix $\mathbf{E}$). Such a strategy is parameterized by the dictionary of steering vectors $\mathbf{E}$. In the case where the dictionary $\mathbf{E}$ is unknown (or imperfectly known), we propose to learn the matrix used in \eqref{eq:estim_strat} directly on data via backpropagation \cite{Rumelhart1986}, using as initialization the matrix $\tilde{\mathbf{E}}$ corresponding to the imperfect physical model.

\begin{figure}[htbp]
\center
\includegraphics[width=0.9\columnwidth]{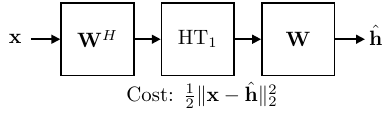}
\caption{Deep unfolding for single path channel estimation.}
\label{fig:mpnet1}
\end{figure}

\noindent{\bf Neural network structure.} Such a neural network structure corresponds to the $k$-sparse autoencoder \cite{Makhzani2013}, which has originally been introduced for image classification. The deep unfolding of channel estimation using a physical model as in \eqref{eq:estim_strat} corresponds to use a $k$-sparse autoencoder, setting the sparsity parameter to $k=1$. This neural network structure is shown on figure~\ref{fig:mpnet1}, where $\text{HT}_1$ refers to the hard thresholding operator which keeps only the entry of greatest modulus of its input and sets all the others to zero. The parameters of this neural network are the weights $\mathbf{W} \in \mathbb{C}^{N \times A}$. Note that complex weights and inputs are handled classically by stacking the real and imaginary parts for vectors and using the real representation for matrices.

\noindent{\bf Training.} The method we propose to jointly estimate channels while simultaneously correcting an imperfect physical model amounts to initialize the network of figure~\ref{fig:mpnet1} with a dictionary of nominal steering vectors $\tilde{\mathbf{E}}$ and then to perform a minibatch gradient descent \cite{Bottou2010} on the cost function $\tfrac{1}{2}\Vert \mathbf{x} - \hat{\mathbf{h}} \Vert_2^2$ to update the weights $\mathbf{W}$ in order to correct the model. It operates \emph{online}, on streaming observations $\mathbf{x}_i,\, i=1,\dots,\infty$ of the form \eqref{eq:observations} acquired over time (coming from all users simultaneously). Note that, as opposed to the classical unfolding strategies \cite{Gregor2010,Hershey2014}, the proposed method is totally \emph{unsupervised}, meaning that it requires only noisy channel observations and no clean channels to run.

\noindent{\bf Implementation details.} 
In all the experiments performed in this letter, we use minibatches of $200$ observations and the Adam optimization algorithm \cite{Kingma2014} with an exponentially decreasing learning rate starting at $0.001$ and being multiplied by $0.9$ every $200$ minibatchs. Moreover, the method was found to perform better with the input data being normalized.  To simplify notation, we denote abusively also $\mathbf{x}_i,\, i=1,\dots,\infty$ the data after normalization.

 \subsection{Generalization to multipath channels}
\begin{figure}[htbp]
\includegraphics[width=\columnwidth]{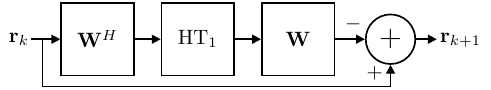}
\caption{{\sf mpNet}: Unfolding for multipath channel estimation.}
\label{fig:mpnetK}
\end{figure}

 Real channels are often not made of a single path, in which case the proposed method becomes suboptimal. Indeed, it uses a $k$-sparse autoencoder with $k=1$, implicitly assuming a single path. However, real world channels are often sparse (well approximated by only a few paths). This is particularly true at millimeter wave frequencies \cite{Samimi2016}. In order to adapt the unfolding strategy to such channels, we propose to apply recursively the structure of figure~\ref{fig:mpnet1}, subtracting at each step the current output from the observation, exactly mimicking the matching pursuit (MP) algorithm \cite{Mallat1993}. The number $K$ of times the structure is replicated (depth of the network) corresponds to the number of estimated paths. 
The neural network corresponding to such a strategy is schematized on figure~\ref{fig:mpnetK}, we call it {\sf mpNet} (for matching pursuit network). It is trained exactly as the network of figure~\ref{fig:mpnet1}, with tied weights across iterations (we tried to untie the weights but observed no improvement) and cost function $\tfrac{1}{2}\Vert \mathbf{x} - \hat{\mathbf{h}} \Vert_2^2 = \tfrac{1}{2}\Vert \mathbf{r}_K \Vert_2^2$.

\begin{figure*}[t]
\begin{subfigure}[b]{0.333\textwidth}
\includegraphics[width=\columnwidth]{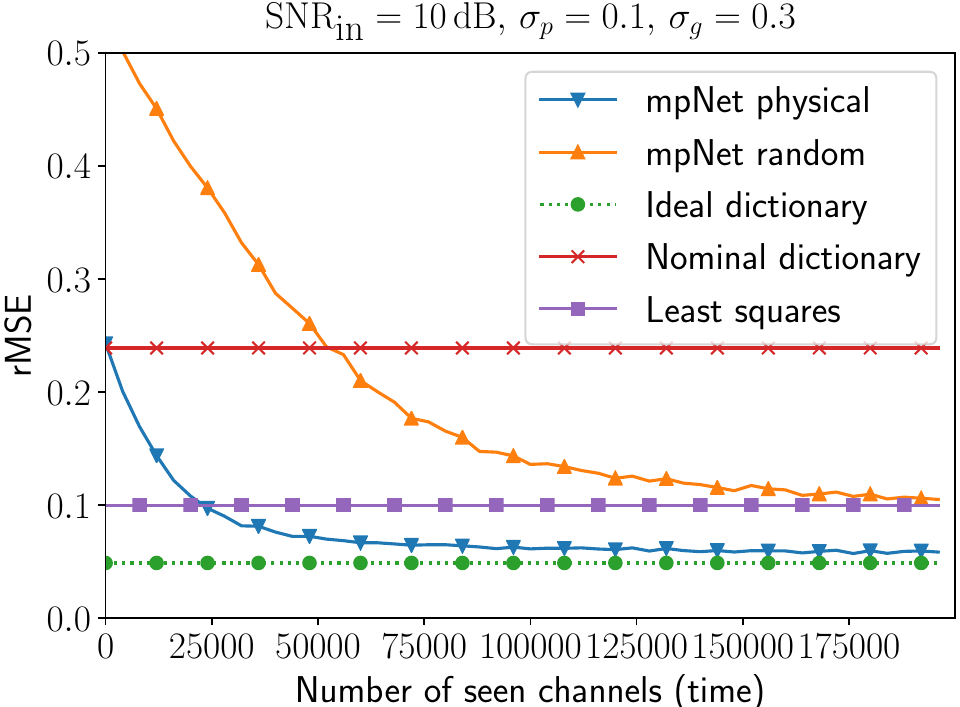}
\end{subfigure}
\begin{subfigure}[b]{0.333\textwidth}
\includegraphics[width=\columnwidth]{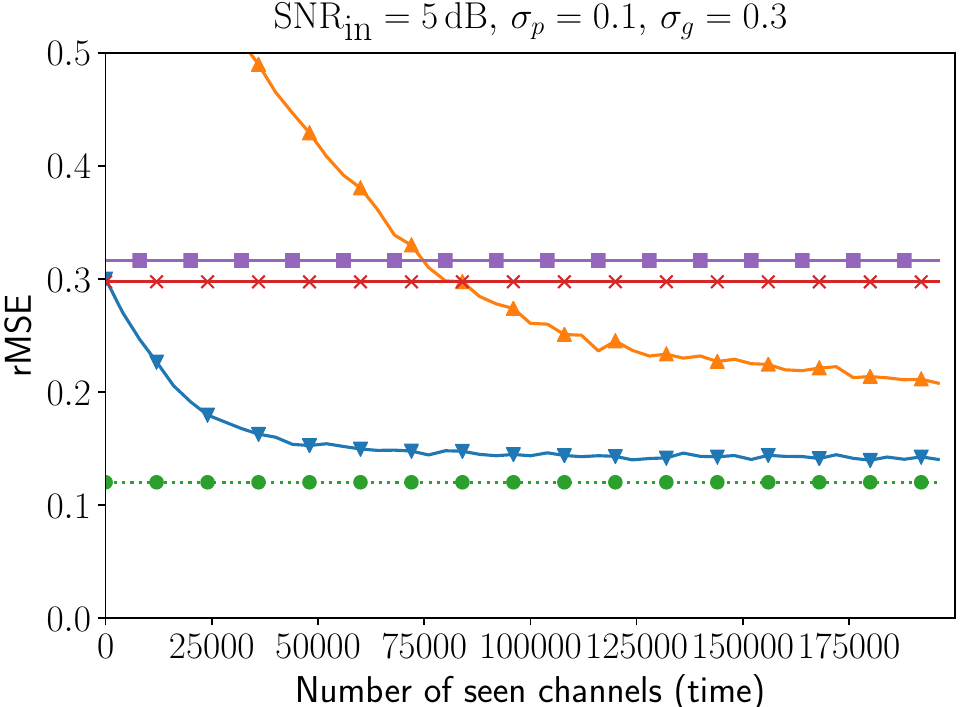}
\end{subfigure}
\begin{subfigure}[b]{0.333\textwidth}
\includegraphics[width=\columnwidth]{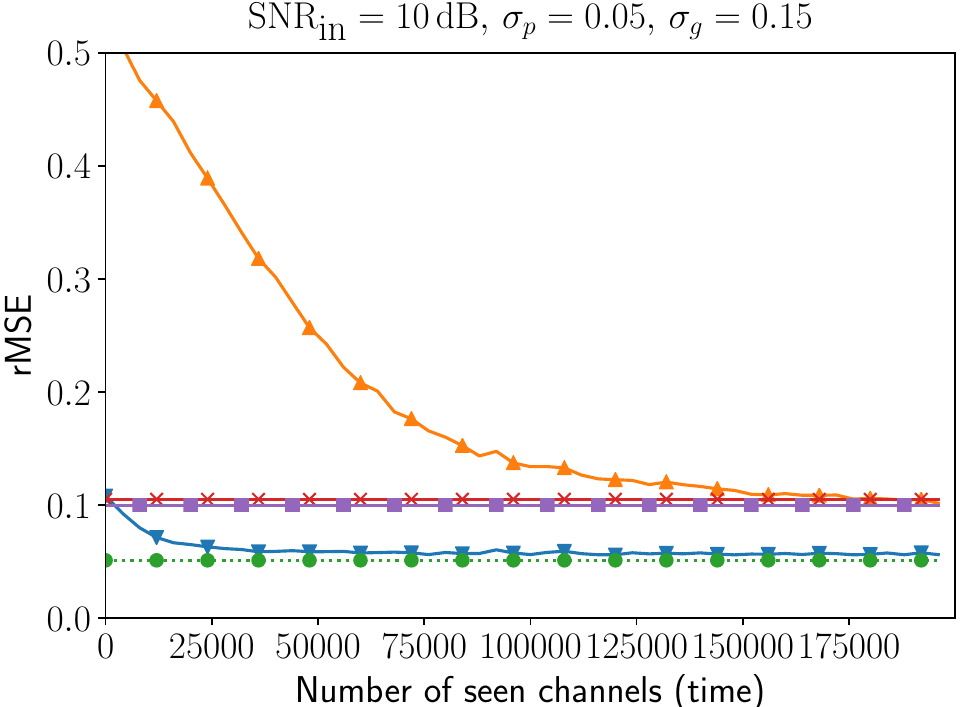}
\end{subfigure}
\caption{Channel estimation performance on synthetic realistic channels for various SNRs and model imperfections.}
\label{fig:multi}
\end{figure*}

\noindent{\bf Computational complexity.} Note that the forward pass in {\sf mpNet} costs $\mathcal{O}(KNA)$ arithmetic operations and the backpropagation step cotst $\mathcal{O}(KN)$ arithmetic operations ($A$ times less). This means that jointly learning the model and estimating the channel (computing the forward and backward passes) is done at a cost that is the same order as the one of simply estimating the channel with a greedy algorithm (MP or OMP) without adapting the model at all to data (which corresponds to computing only the forward pass). This very light computational cost makes the method adapted to online learning, as opposed to previously proposed channel estimation strategies based on deep unfolding \cite{He2018,Wei2019,Vanlier2020}.


\subsection{Experiment}
\noindent{\bf Setting.} Let us now assess {\sf mpNet} on realistic channels. To do so, we consider the SSCM channel model \cite{Samimi2016} in order to generate non-line-of-sight (NLOS) channels at $28\,\text{GHz}$ (see \cite[table~IV]{Samimi2016}) corresponding to all users. We consider the same setting as in section~\ref{sec:impact}, namely a base station equipped with an ULA of $64$ antennas, with an half-wavelength nominal spacing and unit nominal gains used to build the imperfect nominal dictionary $\tilde{\mathbf{E}}$ (with $A=8N$) which serves as an initialization for {\sf mpNet}. The actual antenna arrays are generated the same way as in section~\ref{sec:impact}, using \eqref{eq:imperfection}, and are kept fixed for the whole experiment. We consider two model imperfections: $\sigma_p = 0.05,\,\sigma_g = 0.15$ (small uncertainty) and $\sigma_p = 0.1,\,\sigma_g = 0.3$ (large uncertainty) to build the unknown ideal dictionary $\mathbf{E}$. The input SNR takes the values $\{5,10\}\,\text{dB}$ while the parameter $K$ (controlling the depth of {\sf mpNet}) is set to $\{6,8\}$ respectively (determined by cross validation). The proposed method is compared to the least squares estimator and to the OMP algorithm with $K$ iterations using either the imperfect nominal dictionary $\tilde{\mathbf{E}}$ or the unknown ideal dictionary $\mathbf{E}$. In order to show the interest of the imperfect model initialization, we also compare the proposed method to {\sf mpNet} using a random (Gaussian) initialization. This baseline correspond to a classical online dictionary learning method \cite{Mairal2010}.

\noindent{\bf Results.} The results of this experiment are shown on figure~\ref{fig:multi} as a function of the number of channels of the form~\eqref{eq:observations} seen by the base station over time. The performance measure is the relative mean squared error ($\text{rMSE} = \Vert \hat{\mathbf{h}} - \mathbf{h} \Vert_2^2/\Vert \mathbf{h} \Vert_2^2$) averaged over minibatches of $200$ channels. First of all, the imperfect model is shown to be well corrected by {\sf mpNet}, the blue curve being very close to the green one (ideal unknown dictionary) after a certain amount of time. This is true both for a small uncertainty and for a large one and at all tested SNRs. Note that using the nominal dictionary (initialization of {\sf mpNet}) may be even worse than the least squares method, showing the interest of correcting the model, since with learning {\sf mpNet} always ends up outperforming the least squares. Second, comparing the leftmost and center figures, it is interesting to notice that learning is faster and the attained performance is better with a large SNR (the blue and green curves get closer, faster), which can be explained by the better quality of data used to train the model. Third, comparing the leftmost and rightmost figures, it is apparent that a smaller uncertainty, which means a better initialization since the nominal dictionary is closer to the ideal unknown dictionary, leads to a faster convergence, but obviously also to a smaller improvement. Finally, comparing the blue and orange curves on all figures, it is apparent that initialization matters. Indeed, the random initialization performs much worse than the initialization with the nominal dictionary and takes longer to converge. These conclusions are very promising and highlight the applicability of the proposed method. 
\balance

\section{Conclusion and perspectives}
In this paper, we proposed a method to add flexibility to physical models used for MIMO channel estimation. It is based on the deep unfolding strategy that views classical algorithms as neural networks. The proposed method was shown to correct incrementally (via online learning) an imperfect or imperfectly known physical model in order to make channel estimation as efficient as if the unknown ideal model were known. This claim was empirically validated on realistic millimeter wave outdoor channels, for various SNRs and model imperfections.

We used here uncertainty on the antenna gains and positions to illustrate physical models imperfection, but the presented method applies in principle to any imperfection (be it linear or not). For example, it could correct models in cases where the radiation pattern of the antennas differs from the nominal one, or if the plane wave assumption is not perfectly valid. Moreover, we chose to unfold the MP algorithm, but more sophisticated sparse recovery algorithm could be unfolded the same way (such as approximate message passing \cite{Donoho2009}).


\clearpage
\bibliographystyle{unsrt}
\bibliography{biblio_mimo}

\end{document}